\documentstyle[twocolumn,seceq,epsfig]{jpsj}

\def\runtitle{Product Wave Function Renormalization Group: construction from  
the matrix product point of view}
\def\runauthor{Kouji {\sc Ueda}, Tomotoshi {\sc Nishino},  Kouichi {\sc Okunishi}, 
Yasuhiro {\sc Hieida}, Rene {\sc Derian}, and Andrej {\sc Gendiar}}

\title{Product Wave Function Renormalization Group: construction from  
the matrix product point of view}

\author
{Kouji {\sc Ueda}, Tomotoshi {\sc Nishino}, 
Kouichi {\sc Okunishi},$^{2)}$ Yasuhiro {\sc Hieida},$^{3)}$ \\
Rene {\sc Derian},$^{4)}$ and Andrej {\sc Gendiar}$^{4)}$ }

\inst
{Department of Physics, Faculty of Science, Kobe University, Kobe 657-8501\\
$^{2)}$ Department of Physics, Faculty of Science, Niigata University, Igarashi 2, 
Niigata 950-2181\\
$^{3)}$Computer and Network Center, Saga University, Saga 840-8502 \\
$^{4)}$ Institute of Electrical Engineering, Slovak Academy of Sciences, D\'ubravsk\'a cesta 9, 
SK-841 04 Bratislava, Slovakia}

\abst
{We present a construction of a matrix product state (MPS) that approximates the
largest-eigenvalue eigenvector of a transfer matrix $T$,
for the purpose of rapidly performing the infinite system density matrix
renormalization group (DMRG) method applied to two-dimensional classical lattice 
models. We use the fact that the largest-eigenvalue eigenvector of $T$ can be approximated
by a state vector created from the upper or lower half of a finite size cluster.
Decomposition of the obtained state vector into the MPS 
gives a way of extending the MPS, at the 
system size increment process in the infinite system DMRG algorithm. 
As a result, we successfully give the physical interpretation of 
the product wave function renormalization group (PWFRG) method,
and obtain its appropriate initial condition.
}

\begin{document}
\sloppy
\maketitle

\section{Introduction}

Calculation of the maximum eigenvalue of a transfer matrix is one of the most fundamental 
task in statistical mechanics of lattice models. In two dimensional (2D) classical systems, 
the density matrix renormalization group (DMRG) method is a powerful tool for this purpose,~\cite{DMRG,DMRG2,DMRG3,DMRG4} where the maximum eigenvalue is variationally 
evaluated by implicit use of the matrix product 
state (MPS).~\cite{Fannes,Ostlund,Ostlund2,Ostlund3,Takasaki,Verstraete,Verstraete2,Verstraete3}

In actual applications of the DMRG method to classical models, most of the computational time 
is consumed for the determination of the largest-eigenvalue eigenvector of the renormalized 
transfer matrix.~\cite{DMRG2,DMRG3} The matrix product structure of the variational 
state helpfully tells us how to save the computational time.
For instance, in the {\it finite} system DMRG algorithm~\cite{DMRG4},  a good trial 
vector for the diagonallization of the renormalized transfer matrix
can be constructed from the MPS.~\cite{Acce,Acce2}
For the {\it infinite} system DMRG algorithm, however,  such preparation of the trial 
vector is rather difficult, since we have to extrapolate 
the MPS --- increase the number of matrices --- toward a lager system size. 

The solution of this problem is partially achieved by the product wave function renormalization 
group (PWFRG) method, which increases the number of matrices in the 
MPS on the basis of the relation 
between block-spin transformations.~\cite{PWFRG}
The numerical efficiency of the PWFRG method has been demonstrated for classical 
systems~\cite{Acts,Acts3} as well as quantum 
ones.~\cite{Hieida,Hagiwara,Okunishi,Narumi,Yoshikawa} 
In spite of the efficiency of the PWFRG method, we have 
two problems to be solved from theoretical view point. One is in the
physical interpretation of the extrapolation of the MPS,
in particular, when the system size is small. This is because
the extension process has been justified only in the thermodynamic limit. 
Another is in the preparation of initial matrices, which is necessary to start 
the recursive computation of the PWFRG method. 
So far these matrices are set empirically,
which may not guarantee the numerical stability.

To find the solution of the above problems of the PWFRG method, it is instructive to recall the 
physical substance of the Baxter's corner transfer matrix (CTM) method~\cite{Baxter} 
or its variant, the corner transfer matrix renormalization group (CTMRG) 
method.~\cite{CTMRG,CTMRG2} 
These CTM type methods calculate the partition function of the finite size cluster 
recursively, where the number of iterations corresponds to the linear dimension of
the system. A preferable point on the methods is that the initial condition
is well-defined as CTMs with appropriate boundary spin configurations.
In this paper, we import such 
well-controllable aspects of the CTM-type methods into the extension process of the MPS. 

In the next section we consider a way of the extension of the finite size clusters,
which can be regarded as a variant of the CTMRG method. In \S 3 we analyze the MPS structure 
of the approximate largest-eigenvalue eigenvectors generated by the recursive extension 
process obtained in \S2.
We successfully give the physical interpretation of of the PWFRG method,
and reconstruct its numerical algorithm, which includes the appropriate
initial condition. In the last section, we conclude the obtained results.

\section{Recursive Construction of State Vectors}

As an example of 2D statistical models,
let us consider the interaction-round-a-face (IRF) model, 
which contains the Ising model as its special case. 
 The IRF model is defined by the local Boltzmann Weight 
$W( s'  \sigma' | s \sigma   )$ assigned for each face of the square lattice, where 
$s$, $s'$, $\sigma$ and $\sigma'$ 
represent $q$-state spin variables on lattice points. (See Fig.1.) For simplicity, we consider
2-state Ising spins ($q = 2$) and assume the symmetry 
$W( s'  \sigma' | s \sigma   )$ 
$= W( \sigma' s'  | \sigma s )$  and 
$W( s'  \sigma' | s \sigma   )$ 
$= W( s \sigma  |  s'  \sigma' )$ 
throughout this article.~\cite{general} Let us consider the 
finite size cluster of width $2N$ shown in Fig.1.~\cite{cluster} 
The partition function of the cluster is given by the spin configuration sum
of the product of all the Boltzmann weights in the cluster
\begin{equation}
Z_N^{~} = \sum \prod \, W \, ,
\end{equation}
where we have omitted spin variables for simplicity. 
Throughout this section we assume the 
free boundary condition.

\begin{figure}
\epsfxsize=50mm 
\centerline{\epsffile{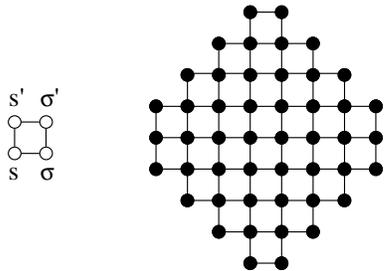}}
\caption{The arrangement of spins of the local Boltzmann weight $W( s'  \sigma' | s \sigma   )$ 
(in the left) and the finite size cluster of width $2N$ considered in the partition function 
$Z_N^{~}$ in Eq.(2.1) (in the right). We show
the case $N = 4$, equivalently, $2N = 8$.}
\label{fig:1}
\end{figure}

We divide the system into the upper  and lower halves.
Figure 2(a) shows the lower half of the system, where we label the spins on the cut as 
$s_1^{~} \, \ldots s_N^{~} \sigma_N \, \ldots \sigma_1$ from the left to right.
Considering the product of local Boltzmann weights in the lower half, and 
taking the configuration sum of spins except for
$s_1^{~} \, \ldots s_N^{~} \sigma_N \, \ldots \sigma_1$,
we obtain a $2^{2N}_{~}$-dimensional vector $| \Psi_N^{~} \rangle$ whose
elements are
 $\Psi_N^{~}[ s_1^{~} \, \ldots s_N^{~} \sigma_N^{~} \, \ldots \sigma_1^{~} ]$.
In the same manner we obtain the $2^{2N}_{~}$-dimensional vector
$\langle \Psi_N^{~} | $ that corresponds to the upper half of the original cluster.
Let us call these vectors {\it state vectors} in the following.
By definitions of $\langle \Psi_N^{~} | $ and $| \Psi_N^{~} \rangle$, 
the partition function in Eq.(2.1) is expressed by the inner product
\begin{equation}
Z_N^{~} = \langle \Psi_N^{~} | \Psi_N^{~} \rangle \, .
\end{equation}
Let us introduce the row-to-row transfer matrix
\begin{eqnarray}
&&T_{N}^{~}[ s'_1 \, \ldots s'_N \sigma'_N \, \ldots \sigma'_1 | 
s_1^{~} \, \ldots s_N^{~} \sigma_N \, \ldots \sigma_1 ] \\
&&= W( s'_1 s'_2 | s_1^{~} s_2^{~} ) 
W( s'_2 s'_3 | s_2^{~} s_3^{~} ) \cdots
W( \sigma'_2 \sigma'_1 | \sigma_2^{~} \sigma_1^{~} )  \, . \nonumber
\end{eqnarray}
For simplicity, we introduce `dot product notation'
with which we abbreviate eq.(2.3) as 
  $T_N^{~} = W \cdot W  \cdots W$. 
  
The state vector $| \Psi_{N}^{~} \rangle$ can be used as a good 
variational vector for $T_{N}^{~}$,
when the system is off critical and $N$ is sufficiently larger than the 
correlation length.~\cite{Baxter} This is because the variational ratio
\begin{equation}
g = \frac{\langle \Psi_{N}^{~} | T_{N}^{~} | \Psi_{N}^{~} \rangle}{
\langle \Psi_{N}^{~} | \Psi_{N}^{~} \rangle} = 
\frac{\langle \Psi_{N}^{~} | T_{N}^{~} | \Psi_{N}^{~} \rangle}{Z_{N}^{~}}
\end{equation}
well approximates the largest-eigenvalue of $T_{N}^{~}$ in the sense that,
as $N$ increases,  $f = - ( k \, \ln \, g ) \, / \, 2N$ where $k$ is the 
Boltzmann constant
 approaches to the free energy per site
 in the thermodynamic limit.
  
\begin{figure}
\epsfxsize=70mm 
\centerline{\epsffile{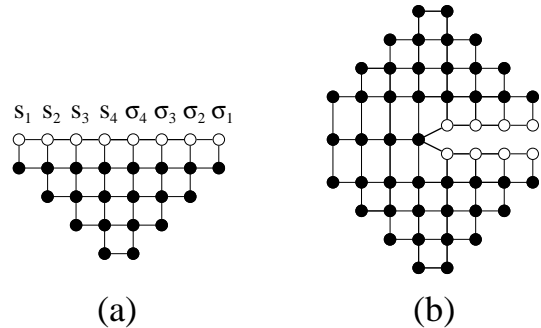}}
\caption{Graphical representations: 
(a) The state vector $| \Psi_N^{~} \rangle$  when $N = 4$. (b) The reduced density matrix 
$\rho_N^{\rm R}$. Black dots denote 
spin variables that are summed.}
\label{fig:2}
\end{figure}

We occasionally interpret  $| \Psi_N^{~} \rangle$ as a 
$2^N_{~} \times 2^N_{~}$ matrix $\Psi_N^{~}$, whose elements
$\Psi_N^{~}( s_N \, \ldots s_1 | \sigma_N^{~} \, \ldots \sigma_1^{~}  )$ 
are equal to 
$\Psi_N^{~}[ s_1^{~} \, \ldots s_N^{~} \sigma_N^{~} \, \ldots \sigma_1^{~} ]$.
 (We have put the vertical line $|$ between the first matrix index
 $s_N^{~} \, \ldots s_1^{~}$ and the second one $\sigma_N^{~} \, \ldots \sigma_1^{~}$,
respectively, which corresponds to the left and the right half-row spins. We 
reverse the spin order as $s_N^{~} \, \ldots s_1^{~}$ for the later convenience.)
This notation is useful when we consider the reduced density matrix
$\rho_N^{\rm R} = {\rm Tr}_{\rm L}^{~} \,  | \Psi_N^{~} \rangle \langle \Psi_N^{~} |$,
where ${\rm Tr}_{\rm L}^{~}$ represents the spin contraction for all the $s_i^{~}$ in the
left side. Figure 2(b) shows the graphical representation of 
$\rho_N^{\rm R}$.~\cite{Baxter,DMRG2,CTMRG}
Using the matrix notation $\Psi_N^{~}$ we can simply write $\rho_N^{\rm R}$
as the matrix product  $\Psi_N^{\dagger} \Psi_N^{~}$, since 
the elements of  $\rho_N^{\rm R}$ are explicitly given by
\begin{eqnarray}
\rho_N^{\rm R}( \sigma'_N \, \ldots \sigma'_1 | \sigma_N^{~} \, \ldots \sigma_1^{~} ) =
&& \sum_{s_N^{~} \, \ldots s_1^{~}}^{~}
\Psi_N^{~}( s_N^{~} \, \ldots s_1^{~} | \sigma'_N \, \ldots \sigma'_1 ) \nonumber\\
&& \Psi_N^{~}( s_N^{~} \, \ldots s_1^{~} | \sigma_N^{~} \, \ldots \sigma_1^{~} ) \, .
\end{eqnarray}
Similarly we obtain the reduced
density matrix for the left side $\rho_N^{\rm L} = {\rm Tr}_{\rm R}^{~} \, | \Psi_N^{~}
\rangle \langle \Psi_N^{~} | = \Psi_N^{~} \Psi_N^{\dagger} $.
It should be noted that all the elements of $\Psi_N^{~}$, $\rho_N^{\rm R}$ and $\rho_N^{\rm L}$ are
real and non negative. The partition function in Eq.(2.2) is expressed as
\begin{equation}
Z_N^{~} = {\rm Tr} \, \rho_N^{\rm L} 
 = {\rm Tr} \, \rho_N^{\rm R}= {\rm Tr} \, \Psi_N^{\dagger} \Psi_N^{~} \, .
\end{equation}

In the following
we consider a recursive construction of $\Psi_N^{~}$, starting from $N = 1$
 up to a certain system size.
Let us first take the configuration sum for 2 spins on the local Boltzmann weight
\begin{equation}
\Psi_1^{~}( s'_1 | \sigma'_1 ) = \sum_{s_1^{~} \sigma_1^{~}}^{~}
W( s'_1 \sigma'_1 | s_1^{~} \sigma_1^{~} ) 
\end{equation}
to obtain a state vector $| \Psi_1^{~} \rangle$ of the smallest width. 
(See Fig.2(a) and imagine the case $N = 1$.) In order to extend the system
size and construct $| \Psi_2^{~} \rangle$, consider the row-to-row transfer matrix
$T_2^{~} \equiv W \cdot W \cdot W$, whose elements are expressed as
\begin{eqnarray}
&& T_2^{~}[ s'_1 s'_2 \sigma'_2 \sigma'_1 | s_1^{~} s_2^{~} \sigma_2^{~} \sigma_1^{~} ]\\
&& ~~~~~~  =  W( s'_1 s'_2 | s_1^{~} s_2^{~} ) 
W( s'_2 \sigma'_2 | s_2^{~} \sigma_2^{~} )
W( \sigma'_2 \sigma'_1 | \sigma_2^{~} \sigma_1^{~} )  \, . \nonumber
\end{eqnarray}
(See Fig.3(a).)
It is convenient to define half-row transfer matrices for both the left and the right parts of
the transfer matrix. For Eq.(2.8), we have
\begin{eqnarray}
T_2^{\rm R} (  \sigma'_2 \sigma'_1 | \sigma_2^{~} \sigma_1^{~} ) &=& 
W( \sigma'_2 \sigma'_1 | \sigma_2^{~} \sigma_1^{~} ) \nonumber\\
T_2^{\rm L} (  s'_2 s'_1 | s_2^{~} s_1^{~} ) &=& W( s'_1 s'_2 | s_1^{~} s_2^{~} ) \, ,
\end{eqnarray}
where the order of spin indices in $T_2^{\rm L}$ is reversed for convenience, so that
$T_2^{\rm L} = T_2^{\rm R}$ holds when $W$ is symmetric. To
obtain $| \Psi_2^{~} \rangle$, we ``join" 
$T_2^{~} \equiv T_2^{\rm L} \cdot W \cdot T_2^{\rm R}$ to $| \Psi_1^{~} \rangle$
as follows.  First we contract a spin in the half-row transfer matrices as
\begin{eqnarray}
P_2^{\rm R} (  \sigma'_2 \sigma'_1 | \sigma_2^{~}  ) &=& \sum_{\sigma_1^{~}}^{~}
T_2^{\rm R} (  \sigma'_2 \sigma'_1 | \sigma_2^{~} \sigma_1^{~} ) \nonumber\\
P_2^{\rm L} (  s'_2 s'_1 |  s_2^{~} ) &=& \sum_{s_1^{~}}^{~} 
T_2^{\rm L} (  s'_2 s'_1 | s_2^{~} s_1^{~} ) \, .
\end{eqnarray}
For the later use we
define a $2^4_{~} \times 2^2_{~}$ matrix 
$P_2^{~} \equiv P_2^{\rm L} \cdot W \cdot P_2^{\rm R}$ 
shown in Fig.3(b), whose elements are given by
\begin{eqnarray}
&& P_2^{~}[ s'_1 s'_2 \sigma'_2 \sigma'_1 | s_2^{~} \sigma_2^{~}  ] \\
&& ~~~~~~  =  
P_2^{\rm L} (  s'_2 s'_1 |  s_2^{~} )
W( s'_2 \sigma'_2 | s_2^{~} \sigma_2^{~} )
P_2^{\rm R} (  \sigma'_2 \sigma'_1 | \sigma_2^{~}  ) \, . \nonumber
\end{eqnarray}
Then we can obtain $| \Psi_2^{~} \rangle$ by applying 
$P_2^{~}$ to $| \Psi_1^{~} \rangle$ as shown in Fig.3(c), where
the matrix elements of $\Psi_2^{~}$ is explicitly given by
\begin{eqnarray}
\Psi_2^{~}( s'_2 s'_1  |  \sigma'_2 \sigma'_1   )
= 
\sum_{s_2^{~} \sigma_2^{~}}^{~} \, 
&& P_2^{\rm L}( s'_2 s'_1  | s_2  ) \\
\times \,  
W( s'_2  \sigma'_2  | s_2^{~} \sigma_2^{~} )
&& P_2^{\rm R}( \sigma'_2 \sigma'_1  | \sigma_2^{~}   ) 
 \Psi_1^{~}( s_2^{~}  |  \sigma_2^{~}) \, .
 \nonumber
\end{eqnarray}
 It should be noted that the variables contained in $ \Psi_1^{~}$ are
$s_2^{~}$ and $\sigma_2^{~}$ only.
In short, we write Eq.(2.12) as
 $| \Psi_2^{~} \rangle = 
 P_2^{\rm L} \cdot W \cdot P_2^{\rm R} \, |  \Psi_1^{~} \rangle$.

\begin{figure}
\epsfxsize=65mm 
\centerline{\epsffile{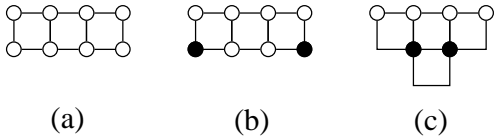}}
\caption{Graphical representations: (a) Transfer matrix $T_2^{~} \equiv W \cdot W \cdot W = T_2^{\rm L} \cdot W \cdot T_2^{\rm R}$ in Eq.(2.8). 
(b) $P_2^{~} \equiv P_2^{\rm L} \cdot W \cdot P_2^{\rm R}$ in Eq.(2.11). 
(c) The extension process of the state vector 
$P_2^{\rm L} \cdot W \cdot P_2^{\rm R} \, |  \Psi_1^{~} \rangle$ in Eq.(2.12).}
\label{fig:3}
\end{figure}

In the same manner we obtain $| \Psi_3^{~} \rangle$ by applying 
$P_3^{~} \equiv P_3^{\rm L} \cdot W \cdot P_3^{\rm R}$ to $| \Psi_2^{~} \rangle$, where the 
matrix elements of $P_3^{\rm R} \equiv W \cdot P_2^{\rm R}$ are given by
\begin{equation}
P_3^{\rm R}( \sigma'_3 \sigma'_2 \sigma'_1 | \sigma_3^{~} \sigma_2^{~} ) 
= W( \sigma'_3 \sigma'_2 | \sigma_3^{~} \sigma_2^{~} ) 
P_2^{\rm R}( \sigma'_2 \sigma'_1 | \sigma_2^{~} ) \, .
\end{equation}
Hereafter we do not refer to the calculations of $T_N^{\rm L}$, $P_N^{\rm L}$,
and $\rho_N^{\rm L}$ for book keeping purpose.

Since such recursive extensions of $| \Psi_N^{~} \rangle$ cause the exponential blow-up
of the vector (or matrix) dimensions, we have to introduce the density matrix 
renormalization to reduce the dimensions below a certain number 
$m$.~\cite{DMRG,DMRG2} The value of $m$ is normally chosen to be $10 \sim 1000$.
(In the following explanation we assume $m \ge 8$.)
Let us consider the case $N = 2$ as an example. We
diagonalize the reduced density matrix
\begin{eqnarray}
\rho_2^{\rm R}( \sigma'_2 \sigma'_1 | \sigma_2^{~} \sigma_1^{~} ) &=&
\sum_{s_2^{~} s_1^{~}}^{~} 
\Psi_2^{~}( s_2^{~} s_1^{~} | \sigma'_2 \sigma'_1  ) 
\Psi_2^{~}( s_2^{~} s_1^{~} | \sigma_2^{~} \sigma_1^{~}  ) \nonumber\\
 =
\sum_{\xi_2^{~}}^{~} && \!\!\!\!\!\!
U_2^{~}( \sigma'_2 \sigma'_1 | \xi_2^{~} ) \,
\lambda_2^{~}( \xi_2^{~} ) \, 
U_2^{~}( \sigma_2^{~} \sigma_1^{~} | \xi_2^{~} ) \, ,
\end{eqnarray}
where we assume the decreasing order for the
eigenvalues $\lambda_2^{~}( \xi_2^{~} )$, which are non-negative. 
As in the DMRG method, the matrix $U_2^{~}$ plays a role of 
the block spin transformation from the group of
2 spins $\sigma_2^{~} \sigma_1^{~}$ to
a new block spin variable $\xi_2^{~}$, which takes 4 states.
By operating $U_2^{\dagger}$ to $P_3^{\rm R}$, we  obtain the renormalized 
matrix $\tilde{P}_3^{\rm R}$ as
\begin{equation}
{\tilde P}_3^{\rm R}(  \sigma'_3 \xi_2^{~}  |  \sigma_3 \sigma_2^{~}  ) = 
\sum_{\sigma'_2 \sigma'_1}^{~} 
U_2^{~}( \sigma'_2 \sigma'_1 | \xi_2^{~} ) 
P_3^{\rm R}( \sigma'_3 \sigma'_2 \sigma'_1 | \sigma_3^{~} \sigma_2^{~}   ) 
\end{equation}
which is graphically represented in Fig.4.
Here we should remark that, in strict sense,  the r.h.s. of Eq.(2.15) is not a simple matrix multiplication
between ${U}_2^{\dagger}$ and $P_3^{\rm R}$, 
since $U_2^{\dagger}$ works only on $ \sigma'_2$ and $\sigma'_1$.
Keeping this point in mind, we compactly write Eq.(2.15) as
${\tilde P}_3^{\rm R} = {\hat U}_2^{\dagger} P_3^{\rm R}$.
In the following,  we shall put ``hat'' symbol on matrices that play a role 
of the block spin transformations and  ``tilde'' 
on vectors and matrices generated by the block spin transformations. 
Applying ${\tilde P}_3^{~} \equiv {\tilde P}_3^{\rm L} \cdot W \cdot {\tilde P}_3^{\rm R}$
to $| \Psi_2^{~} \rangle$ we obtain an extended state vector
\begin{eqnarray}
{\tilde \Psi}_3^{~}( s'_3 \zeta_2^{~}  |  \sigma'_3 \xi_2^{~}   )
= 
\sum_{s_3^{~} s_2^{~} \sigma_3^{~} \sigma_2^{~} }^{~}
&&{\tilde P}_3^{\rm L}( s'_3\zeta_2^{~}  | s_3^{~} s_2  ) \\
\times \,  W( s'_3  \sigma'_3  | s_3^{~} \sigma_3^{~} )
&& {\tilde P}_3^{\rm R}( \sigma'_3 \xi_2^{~}  | \sigma_3 \sigma_2^{~}   ) 
 \Psi_2^{~}( s_3^{~} s_2^{~}  |  \sigma_3^{~} \sigma_2^{~}) \, ,
 \nonumber
\end{eqnarray}
which contains block spin variables as shown in Fig.5(a). 
Since  the block-spin transformation is exact,
 the relation $Z_3^{~} = \langle \Psi_3^{~} | \Psi_3^{~} \rangle
= \langle {\tilde \Psi}_3^{~} | {\tilde \Psi}_3^{~} \rangle$ is satisfied.

\begin{figure}
\epsfxsize=30mm 
\centerline{\epsffile{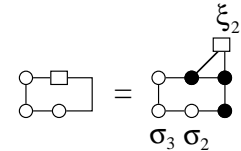}}
\caption{The block spin transformation
${\tilde P}_3^{\rm R} = {\hat U}_2^{\dagger} P_3^{\rm R}$ in Eq.(2.15).}
\label{fig:4}
\end{figure}

Let us continue the system extension process.
We diagonalize the reduced density matrix
\begin{eqnarray}
{\tilde \rho}_3^{\rm R}( \sigma'_3 \xi'_2  |  \sigma_3^{~} \xi_2^{~}  ) &=& 
\sum_{s_3^{~} \zeta_2^{~} }^{~} 
{\tilde \Psi}_3^{~}( s_3^{~} \zeta_2^{~}  |  \sigma'_3 \xi'_2   )
{\tilde \Psi}_3^{~}( s_3^{~} \zeta_2^{~}  |  \sigma_3^{~} \xi_2^{~}   ) \nonumber\\
=
\sum_{\xi_3^{~} }^{~} && \!\!\!\!\!\!
U_3^{~}( \sigma'_3 \xi'_2  | \xi_3^{~} ) \,
\lambda_3^{~}( \xi_3^{~} ) \, 
U_3^{~}( \sigma_3^{~} \xi_2^{~} | \xi_3^{~} ) \, 
\end{eqnarray}
in order 
to obtain the block spin transformation ${\hat U}_3^{~}$ that maps
$\sigma_3^{~} \xi_2^{~}$ to the 8-state block spin $\xi_3^{~}$. 
This time we perform the extension from ${\tilde P}_3^{\rm R}$ to 
${\tilde P}_4^{\rm R}$ operating ${\hat U}_3^{\dagger}$ from the left 
and ${\hat U}_2^{~}$ from the right 
\begin{eqnarray}
{\tilde P}_4^{\rm R}(  \sigma'_4 \xi_3^{~}  |  \sigma_4 \eta_3^{~}  ) \,\, && = 
 \sum_{\sigma'_3 \xi^{~}_2 \sigma_3^{~} \sigma_2^{~}}^{~} 
 U_3^{~}( \sigma'_3 \xi^{~}_2 | \xi_3^{~} ) 
W( \sigma'_4 \sigma'_3 | \sigma_4^{~} \sigma_3^{~} ) \nonumber\\
&& \times \, {\tilde P}_3^{\rm R}(  \sigma'_3 \xi^{~}_2 |  \sigma_3^{~} \sigma_2^{~}  ) \, 
U_2^{~}( \sigma_3^{~} \sigma_2^{~} | \eta_3^{~} ) 
\end{eqnarray}
as shown in Fig.5(b), or in short
${\tilde P}_4^{\rm R} \equiv {\hat U}_3^{\dagger} ( W \cdot {\tilde P}_3^{\rm R} ) {\hat U}_2^{~}$. 
(Note that the matrix  $U_2^{~}$ in the right hand side contains 
$\sigma_3^{~}$ and $\sigma_2^{~}$, not 
$\sigma_2^{~}$ and $\sigma_1^{~}$.)
Up to this stage, we keep all the block spin states since we have
assumed $m \ge 8$. 

\begin{figure}
\epsfxsize=60mm 
\centerline{\epsffile{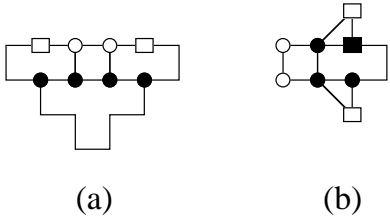}}
\caption{Graphical representations: (a) Extension from $| \Psi_2^{~} \rangle$ to 
$| {\tilde \Psi}_3^{~} \rangle = {\tilde P}_3^{~} | \Psi_2^{~} \rangle$ in Eq.(2.16). (b)
Extension from ${\tilde P}_3^{\rm R}$ to 
${\tilde P}_4^{\rm R} \equiv {\hat U}_3^{\dagger} ( W \cdot {\tilde P}_3^{\rm R} ) {\hat U}_2^{~}$
in Eq.(2.18).}
\label{fig:5}
\end{figure}

Now we can generalize the extension process up to arbitrary system size. If we have
$| {\tilde \Psi}_i^{~} \rangle$ up to a certain size $i \ge 2$, we can construct the extended one
\begin{equation}
| {\tilde \Psi}_{i+1}^{~} \rangle = {\tilde P}_{i+1}^{~} | {\tilde \Psi}_{i}^{~} \rangle
= {\tilde P}_{i+1}^{\rm L} 
\cdot W \cdot {\tilde P}_{i+1}^{\rm R} \, 
| {\tilde \Psi}_{i}^{~} \rangle 
\end{equation}
with
\begin{equation}
{\tilde P}_{i + 1}^{\rm R} \equiv {\hat U}_i^{\dagger} ( W \cdot {\tilde P}_i^{\rm R} ) 
{\hat U}_{i-1}^{~}  \, ,
\end{equation}
where
the block spin transformations ${\hat U}_i^{~}$ and  ${\hat U}_{i-1}^{~}$ 
have been obtained by the 
diagonalizations
of ${\tilde \rho}_i^{\rm R}$ and ${\tilde \rho}_{i-1}^{\rm R}$, respectively. 
The matrix elements of Eq.(2.19) and (2.20) are explicitly given by 
\begin{eqnarray}
{\tilde \Psi}_{i+1}^{~}( s'_{i+1} \zeta_{i}^{~}  && |  \sigma'_{i+1} \xi_{i}^{~}   )
= \nonumber\\
 \sum_{s_{i+1}^{~} \mu_i^{~} \sigma_{i+1}^{~} \eta_{i}^{~}}^{~} 
&&{\tilde P}_{i+1}^{\rm L}( s'_{i+1}\zeta_i^{~}  | s_{i+1}^{~} \mu_i^{~}  )  
W( s'_{i+1}  \sigma'_{i+1}  | s_{i+1}^{~} \sigma_{i+1}^{~} ) \nonumber\\
 && \!\!\!\!\!\!\!\!\!\!\!\!  \times \,
   {\tilde P}_{i+1}^{\rm R}( \sigma'_{i+1} \xi_i^{~}  | \sigma_{i+1} \eta_i^{~}   )
{\tilde \Psi}_i^{~}( s_{i+1}^{~} \mu_i^{~}  |  \sigma_{i+1}^{~} \eta_i^{~}) 
\end{eqnarray}
and
\begin{eqnarray}
&&{\tilde P}_{i+1}^{\rm R}(  \sigma'_{i+1} \xi_i^{~}  |  \sigma_{i+1} \eta_i^{~}  ) = 
 \sum_{\sigma'_i \xi^{~}_{i-1} \sigma_i^{~} \eta_{i-1}^{~}}^{~} 
 U_i^{~}( \sigma'_i \xi^{~}_{i-1} | \xi_i^{~} ) \nonumber\\
&&\times \,
W( \sigma'_{i+1} \sigma'_i | \sigma_{i+1}^{~} \sigma_i^{~} )
  {\tilde P}_i^{\rm R}(  \sigma'_i \xi^{~}_{i-1} |  \sigma_i^{~} \eta_{i-1}^{~}  ) 
 U_{i-1}^{~}( \sigma_i^{~} \eta_{i-1}^{~} | \eta_i^{~} )  \, . \nonumber\\
\end{eqnarray}
The key point in Eqs.(2.20) and (2.22) is that the label
of $U_i^{~}$ is larger than that of  $U_{i-1}^{~}$ by one, 
which enables us to "extrapolate" the 
state vector in Eqs.(2.19) and (2.21).

As is performed in the
DMRG method, we keep at most $m$ numbers of relevant states for the block spin 
variables, and neglect the rest of irrelevant ones in  Eq.(2.20). 
We can thus calculate the approximate partition function
${\tilde Z}_i^{~} = \langle {\tilde \Psi}_{i}^{~} | {\tilde \Psi}_{i}^{~} \rangle$
recursively up to arbitrary system size $i = N$. 
In addition to ${\tilde Z}_N^{~}$, we can
obtain the nearest neighbor spin correlation function
\begin{equation}
\langle s_N^{~} \sigma_N^{~} \rangle 
= \frac{\langle {\tilde \Psi}_N^{~} | s_N^{~} \sigma_N^{~} | {\tilde \Psi}_N^{~} \rangle}{
\langle {\tilde \Psi}_N^{~} | {\tilde \Psi}_N^{~} \rangle}
\end{equation}
at the center of the system.  

\begin{figure}
\epsfxsize=65mm 
\centerline{\epsffile{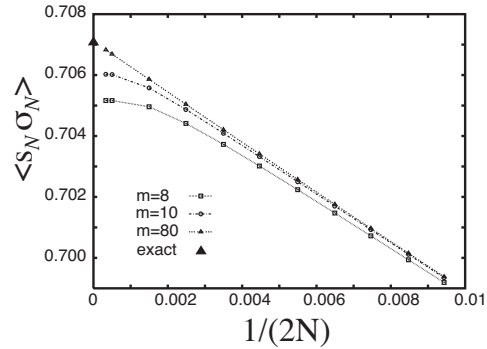}}
\caption{Nearest neighbor spin correlation function $\langle s_N^{~} \sigma_N^{~} \rangle$ 
in Eq.(2.23) of the Ising model at criticality.}
\label{fig:a}
\end{figure}

Let us observe the numerical efficiency of the explained recursive algorithm
in the application to the Ising model.
Figure 6 shows the calculated $\langle s_N^{~} \sigma_N^{~} \rangle $ 
at criticality where the correlation length is infinite. 
The calculated data deviates from the scaling line when the system size exceeds
the artificial cut-off length introduced by the restriction of the degrees of freedom
down to $m$.
The observed cut-off effect
is similar to that of the CTMRG method.~\cite{CTMRG,CTMRG2}
This is because the system-size extension process defined by Eq.(2.19) and (2.20) is quite similar to  
the extension of the corner transfer matrix in the CTMRG method.

\section{Product Wave Function Renormalization Group}

It is worth looking at the structure of 
${\tilde P}_i^{~} \equiv {\tilde P}_i^{\rm L} \cdot W \cdot {\tilde P}_i^{\rm R}$ 
from the view point of the MPS.~\cite{Ostlund,Ostlund2,Ostlund3} Tracing back the
recursive construction in Eqs.(2.10), (2.15), (2.18), and (2.20), we can represent
${\tilde P}_i^{\rm R}( \sigma'_{i} \xi_{i-1}^{~} | \sigma_{i}^{~} \eta_{i-1}^{~} )$ for $i \geq 6$ as
\begin{eqnarray}
{\tilde P}_i^{\rm R} 
&\equiv& {\hat U}_{i-1}^{\dagger} ( W \cdot {\tilde P}_{i-1}^{\rm R} ) 
{\hat U}_{i-2}^{~} \nonumber\\
&=&
{\hat U}_{i-1}^{\dagger} ( W \cdot {\hat U}_{i-2}^{\dagger} ( W \cdot {\tilde P}_{i-2}^{\rm R} ) 
{\hat U}_{i-3}^{~} ) 
{\hat U}_{i-2}^{~} \\
&=&
 {\hat U}_{i-1}^{\dagger} ( W \cdot 
{\hat U}_{i-2}^{\dagger} ( W \cdot ( \cdots ( W \cdot {P}_2^{\rm R} ) \cdots ) 
{\hat U}_{i-3}^{~}
 ) {\hat U}_{i-2}^{~} \, . \nonumber
\end{eqnarray}
Introducing the $2^i_{~}$ by $2^{i-1}_{~}$ matrix 
$P_i^{\rm R}( \sigma'_i \ldots \sigma'_1 | \sigma_i^{~} \ldots \sigma_2^{~} )$, which
is defined by $P_i^{\rm R} \equiv W \cdot P_{i-1}^{\rm R} = 
(W \cdot )^{i-2} P_{2}^{\rm R}$, we rewrite Eq.(3.1) as the simpler form
\begin{equation}
{\tilde P}_i^{\rm R}  \equiv
{\hat U}_{i-1}^{\dagger} \cdots  {\hat U}_3^{\dagger} {\hat U}_2^{\dagger} 
\,\,\,\, P_i^{\rm R} \,\,\,\, 
{\hat U}_2^{~} {\hat U}_3^{~} \cdots {\hat U}_{i-2}^{~}  \, ,
\end{equation}
where  ${\hat U}_{i-1}^{\dagger} \cdots  {\hat U}_3^{\dagger} {\hat U}_2^{\dagger}$
represents the successive block spin transformations by 
$U_2^{~}( \sigma'_2 \sigma'_1 | \xi_2^{~} )$, 
$U_3^{~}( \sigma'_3 \xi_2^{~} | \xi_3^{~} )$\ldots, and
$U_{i-1}^{~}( \sigma'_{i-1} \xi_{i-2}^{~} | \xi_{i-1}^{~} )$.
In the same manner ${\hat U}_2^{~} {\hat U}_3^{~} \cdots {\hat U}_{i-2}^{~}$ represents 
the successive block spin transformations by 
$U_2^{~}( \sigma_3^{~} \sigma_2^{~} | \eta_3^{~} )$, 
$U_3^{~}( \sigma_4^{~} \eta_3^{~} | \eta_4^{~} )$\ldots, and
$U_{i-2}^{~}( \sigma_{i-1}^{~} \eta_{i-2}^{~} | \eta_{i-1}^{~} )$, where one 
should pay attention to the indices of spin variables.
The structure of ${\tilde P}_i^{\rm R}$ is similar to that of the
renormalized half-row transfer matrix
\begin{eqnarray}
{\tilde T}_i^{\rm R} 
&\equiv& {\hat U}_{i-1}^{\dagger} ( W \cdot {\tilde T}_{i-1}^{\rm R} ) {\hat U}_{i-1}^{~}  \nonumber\\
&=& 
{\hat U}_{i-1}^{\dagger} ( W \cdot 
{\hat U}_{i-2}^{\dagger} ( W \cdot ( \cdots ( W \cdot {T}_2^{\rm R} ) \cdots ) {\hat U}_{i-2}^{~}
 ) {\hat U}_{i-1}^{~} \nonumber\\
  &=&
 {\hat U}_{i-1}^{\dagger} \cdots  {\hat U}_3^{\dagger} {\hat U}_2^{\dagger} 
\,\,\,\, ( W \cdots W \cdot T_2^{\rm R} ) \,\,\,\, 
 {\hat U}_2^{~} {\hat U}_3^{~} \cdots {\hat U}_{i-1}^{~} \nonumber\\
 &=&
 {\hat U}_{i-1}^{\dagger} \cdots  {\hat U}_3^{\dagger} {\hat U}_2^{\dagger} 
\,\,\,\, T_i^{\rm R} \,\,\,\, 
 {\hat U}_2^{~} {\hat U}_3^{~} \cdots {\hat U}_{i-1}^{~}  \, ,
\end{eqnarray}
which has been used in the DMRG method. 
(We have introduced $T_i^{\rm R} = W \cdot T_{i-1}^{\rm R} = ( W \cdot )^{i-2}_{~} T_2^{\rm R}$.)
In Eq.(3.3) the successive block spin transformations
$ {\hat U}_{i-1}^{\dagger} \cdots  {\hat U}_3^{\dagger} {\hat U}_2^{\dagger}$
are the conjugate of
$ {\hat U}_{i-1}^{~} \cdots  {\hat U}_3^{~} {\hat U}_2^{~}$,
and are performed by the matrices
$U_2^{~}( \sigma_2^{~} \sigma_1^{~} | \xi_2^{~} )$, 
$U_3^{~}( \sigma_3^{~} \xi_2^{~} | \xi_3^{~} )$\ldots, and
$U_{i-1}^{~}( \sigma_{i-1}^{~} \xi_{i-2}^{~} | \xi_{i-1}^{~} )$.
Comparing Equations (3.2) and (3.3), let us consider the relation between the renormalization
process explained in the previous section and that in the conventional DMRG method.

Since the structure of
${\tilde T}_i^{\rm R} $ in Eq.(3.3) is similar to that of  ${\tilde P}_i^{\rm R} $
in Eq.(3.2),  we try to handle
${\tilde P}_i^{\rm R} $ and ${\tilde T}_i^{\rm R}$ in a unified manner.
Let us introduce a new matrix $V_1^{~}$, whose elements are given by
\begin{equation}
V_1^{~}( \sigma_2^{~} \sigma_1^{~} | \sigma'_2 ) = \delta( \sigma_2^{~} | \sigma'_2 ) \, ,
\end{equation}
which is independent of $\sigma_1^{~}$, and the right hand side is the
Kronecker delta.
The matrix $V_1^{~}$ has a function of taking configuration 
sum for $\sigma_1^{~}$ as Eq.(2.10), when it is 
operated to the half-row transfer matrices. 
For example, we obtain
\begin{eqnarray}
&& P_i^{\rm R}( \sigma'_i \ldots \sigma'_1 | \sigma_i^{~} \ldots \sigma_2^{~} ) \nonumber\\
&& = \sum_{\sigma''_2 \sigma''_1}^{~} \, 
T_i^{\rm R}( \sigma'_i \ldots \sigma'_1 | \sigma_i^{~} \ldots \sigma_3^{~} \sigma''_2 \sigma''_1 )  \,
V_1^{~}( \sigma''_2 \sigma''_1 | \sigma_2 ) \nonumber\\
&& = \sum_{\sigma_1^{~} }^{~} \, 
T_i^{\rm R}( \sigma'_i \ldots \sigma'_1 | \sigma_i^{~} \ldots \sigma_2^{~} \sigma_1^{~} ) \, ,
\end{eqnarray}
which we compactly write as $P_i^{\rm R}  = T_i^{\rm R}  \, {\hat V}_1^{~}$.
Then we can rewrite ${\tilde P}_i^{\rm R}$ in Eq.(3.2) as
\begin{equation}
{\tilde P}_i^{\rm R} = 
{\hat U}_{i-1}^{\dagger} \cdots  {\hat U}_3^{\dagger} {\hat U}_2^{\dagger} 
\,\,\,\, T_i^{\rm R} \,\,\,\,
{\hat V}_1^{~} {\hat U}_2^{~} {\hat U}_3^{~} \cdots {\hat U}_{i-2}^{~} \, .
\end{equation}
Note  that $V_1^{~}$ in Eq.(3.4) represents the free boundary condition 
of the system under consideration. For the system with fixed boundary condition
$\sigma_1^{~} = 1$, the r.h.s of Eq.(3.4) should be replaced by 
$\delta( \sigma_2^{~} | \sigma'_2 ) \delta( \sigma_1^{~} | 1 )$.

For the moment let us consider the ideal case $m = 2^{i-1}_{~}$,
where we keep all the states for every block spin 
transformations. In such a case the block spin transformation is exact, and  the relation
\begin{eqnarray}
\left[ {\hat U}_2^{~} {\hat U}_3^{~} \cdots {\hat U}_{i-1}^{~} \right]
\left[ {\hat U}_2^{~} {\hat U}_3^{~} \cdots {\hat U}_{i-1}^{~} \right]^{\dagger}_{~} &=& \\
\left[ {\hat U}_2^{~} {\hat U}_3^{~} \cdots {\hat U}_{i-1}^{~} \right]
\left[ {\hat U}_{i-1}^{\dagger} \cdots  {\hat U}_3^{\dagger} {\hat U}_2^{\dagger} \right] 
&=&  I_{i-1}^{~} \nonumber
\end{eqnarray}
is satisfied for the transformation matrices
$U_2^{~}( \sigma_2^{~} \sigma_1^{~} | \xi_2^{~} )$, 
$U_3^{~}( \sigma_3^{~} \xi_2^{~} | \xi_3^{~} )$\ldots, and
$U_{i-1}^{~}( \sigma_{i-1}^{~} \xi_{i-2}^{~} | \xi_{i-1}^{~} )$. The identity
$I_{i-1}^{~}$ 
represents $\delta( \sigma'_1 | \sigma_1^{~} ) 
\delta( \sigma'_2 | \sigma_2^{~} )\ldots$
$\delta( \sigma'_{i-1} | \sigma_{i-1}^{~} )$, and it satisfies 
$I_{i-1}^{~} = id \cdot I_{i-2}^{~} = ( id \cdot )^{i-1}_{~}$ where $id$ is the
local identity $id( \sigma'_{~} | \sigma ) = \delta( \sigma'_{~} | \sigma )$. 
Putting the identity $I_{i}^{~} = id \cdot I_{i-1}^{~}$ after $T_i^{\rm R}$ in Eq.(3.6)
and substituting Eq.(3.7), we obtain 
\begin{eqnarray}
{\tilde P}_i^{\rm R} = && \,\,
{\hat U}_{i-1}^{\dagger} \cdots  {\hat U}_3^{\dagger} {\hat U}_2^{\dagger} 
\,\,\,\, T_i^{\rm R} \,\,\,\,
(  I_{i}^{~} )^2_{~} \,\,\,\,
{\hat V}_1^{~} {\hat U}_2^{~} {\hat U}_3^{~} \cdots {\hat U}_{i-2}^{~} \nonumber\\
= && \,\,
\left[ {\hat U}_{i-1}^{\dagger} \cdots  {\hat U}_3^{\dagger} {\hat U}_2^{\dagger} 
\,\,\,\, T_i^{\rm R} \,\,\,\,
 {\hat U}_2^{~} {\hat U}_3^{~} \cdots {\hat U}_{i-1}^{~}  \right] \\
\times && \,\,
\left[ {\hat U}_{i-1}^{\dagger} \cdots  {\hat U}_3^{\dagger} {\hat U}_2^{\dagger} 
 \,\,\,\,  I_{i}^{~}  \,\,\,\,
{\hat V}_1^{~} {\hat U}_2^{~} {\hat U}_3^{~} \cdots {\hat U}_{i-2}^{~} \right] \, . \nonumber
\end{eqnarray}
For convenience, let us define a matrix
\begin{eqnarray}
{\tilde E}_i^{\rm R} &=&
 {\hat U}_{i-1}^{\dagger} \cdots  {\hat U}_3^{\dagger} {\hat U}_2^{\dagger} 
 \,\,\,\,  I_{i}^{~}  \,\,\,\,
{\hat V}_1^{~} {\hat U}_2^{~} {\hat U}_3^{~} \cdots {\hat U}_{i-2}^{~} \\
&=& id \cdot \left(
 {\hat U}_{i-1}^{\dagger} \cdots  {\hat U}_3^{\dagger} {\hat U}_2^{\dagger} 
 \,\,\,\, I_{i-1}^{~} \,\,\,\,
 {\hat V}_1^{~} {\hat U}_2^{~} {\hat U}_3^{~} \cdots {\hat U}_{i-2}^{~}  \right) \, ,\nonumber
\end{eqnarray}
which satisfies 
$
{\tilde E}_i^{\rm R}  = 
{\hat U}_{i-1}^{\dagger}  \left( id \cdot {\tilde E}_{i-1}^{\rm R} \right)  {\hat U}_{i-2}^{~} 
= id \cdot \left(  {\hat U}_{i-1}^{\dagger}   {\tilde E}_{i-1}^{\rm R}   {\hat U}_{i-2}^{~} \right)
$, or equivalently
\begin{eqnarray}
&& {\tilde E}_i^{\rm R}( \sigma'_i \xi_{i-1}^{~} | \sigma_i^{~} \eta_{i-1}^{~} ) =  
\delta( \sigma'_i | \sigma_i^{~} ) 
\sum_{\sigma_{i-1}^{~} \sigma'_{i-1} \xi_{i-2}^{~}  \eta_{i-2}^{~}}^{~} \nonumber\\
&& \,\,\,\,\,\,  U_{i-1}^{~}( \sigma'_{i-1} \xi_{i-2}^{~} | \xi_{i-1}^{~} ) \,
 {\tilde E}_{i-1}^{\rm R}( \sigma'_{i-1} \xi_{i-2}^{~} | \sigma_{i-1}^{~} \eta_{i-2}^{~} ) \nonumber\\
&& \,\,\,\,\,\, \times \, U_{i-2}^{~}( \sigma_{i-1}^{~} \eta_{i-2}^{~} | \eta_{i-1}^{~} ) \, .
\end{eqnarray}
By definition the element 
${\tilde E}_i^{\rm R}( \sigma'_i \xi_{i-1}^{~} | \sigma_i^{~} \eta_{i-1}^{~} )$ 
is zero if $\sigma'_i \neq  \sigma_i^{~}$.
In the same manner we define ${\tilde E}_i^{\rm L}$, 
that is the same as ${\tilde E}_i^{\rm R}$
by the assumed symmetry of the system. Substituting Eqs.(3.3) and (3.9) to Eq.(3.8) we obtain
the simple relation
\begin{equation}
{\tilde P}_i^{\rm R} = {\tilde T}_i^{\rm R}  \, {\tilde E}_i^{\rm R} 
\end{equation}
among the matrices
${\tilde P}_i^{\rm R}( \sigma'_i \xi_{i-1}^{~} | \sigma_i^{~} \eta_{i-1}^{~} )$,
${\tilde T}_i^{\rm R}( \sigma'_i \xi_{i-1}^{~} | \sigma''_i \xi'_{i-1} )$, and
${\tilde E}_i^{\rm R} ( \sigma''_i \xi'_{i-1} | \sigma_i^{~} \eta_{i-1}^{~} )$.
 Introducing the notations
${\tilde E}_i^{~} \equiv {\tilde E}_i^{\rm L} \cdot {\tilde E}_i^{\rm R}$ and 
${\tilde T}_i^{~} \equiv {\tilde T}_i^{\rm L} \cdot W \cdot {\tilde T}_i^{\rm R}$
we finally reach the relation
\begin{equation}
{\tilde P}_i^{~} = {\tilde T}_i^{~} {\tilde E}_i^{~}
\end{equation}
 for the ideal case 
 where there is no cut-off in the block spin transformation.

\begin{figure}
\epsfxsize=55mm 
\centerline{\epsffile{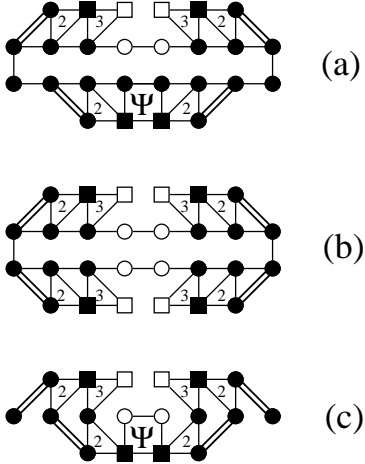}}
\caption{Graphical representations: (a) ${\tilde P}_i^{~} | {\tilde \Psi}_{i-1}^{~} \rangle$ in 
Eq.(2.19)' in case of 
$i = 4$. (b) 
${\tilde T}_i^{~} \equiv {\tilde T}_i^{\rm L} \cdot W \cdot {\tilde T}_i^{\rm R}$. (c)
${\tilde E}_i^{~} | {\tilde \Psi}_{i-1}^{~} \rangle$ in Eq.(3.14). The long rectangular in (a) and (b) 
represents $T_i^{~}$ in Eq.(2.3). The triangle with number $j$ represents 
the block spin transformation ${\hat U}_j^{~}$ or $\hat{U}_j^\dagger$. The rectangular area labeled by 
``$\Psi$'' represents $| {\tilde \Psi}_{i-1}^{~} \rangle$.
Small circles and squares are original spins 
and block spins, respectively, where black ones are summed out. Two circles that are
connected by the double lines represent the same spin. We have used Eq.(3.4), and
therefore  ${\hat V}_1^{~}$ is not explicitly drawn in the picture. 
Stacking (b) and (c) we obtain (a) when Eq.(3.7) is satisfied.}
\label{fig:x}
\end{figure}

In realistic numerical calculation we have to neglect irrelevant states, therefore Eqs.(3.7)-(3.12)
 do not hold exactly, but still ${\tilde P}_i^{~}$ is well approximated
by  ${\tilde T}_i^{~} {\tilde E}_i^{~}$ if sufficiently large numbers of the block spin
states are kept. Under this situation let us consider the operation of ${\tilde P}_i^{~}$
to the state vector
\begin{eqnarray}
| {\tilde \Psi}_{i}^{~} \rangle
= && \,\, {\tilde P}_i^{~} | {\tilde \Psi}_{i - 1}^{~} \rangle ~~~~~~~~~~~~~~~~~~~~~~~{(2.19)'}\nonumber\\
\sim && \,\, {\tilde T}_i^{~} {\tilde E}_i^{~} | {\tilde \Psi}_{i - 1}^{~} \rangle
= {\tilde T}_i^{~} | {\tilde \Phi}_{i }^{~} \rangle 
\end{eqnarray}
as shown in Fig.7(a), where the whole structure of ${\tilde T}_i^{~}$ is shown in 
Fig.7(b). We have introduced a new state vector 
\begin{equation}
 | {\tilde \Phi}_{i }^{~} \rangle 
= {\tilde E}_i^{~} | {\tilde \Psi}_{i - 1}^{~} \rangle 
= {\tilde E}_i^{\rm L} \cdot {\tilde E}_i^{\rm R } | {\tilde \Psi}_{i - 1}^{~} \rangle \, ,
\end{equation}
whose structure is shown in Fig.7(c).
If we use the matrix representation for the 
state vectors  $| {\tilde \Phi}_{i }^{~} \rangle $ and
$| {\tilde \Psi}_{i - 1}^{~} \rangle$,
we can rewrite the above equation by the matrix product 
\begin{equation}
{\tilde \Phi}_{i }^{~} = 
\left( {\tilde E}_i^{\rm L} \right) 
{\tilde \Psi}_{i - 1}^{~}
\left( {\tilde E}_i^{\rm R} \right)^{\dagger}_{~} 
\end{equation}
among  
${\tilde E}_i^{\rm L}( s'_i \zeta'_{i-1} | s_i^{~} \zeta_{i-1}^{~} )$,
${\tilde \Psi}_{i - 1}^{~}( s_i^{~} \zeta_{i-1}^{~} | \sigma_i^{~} \xi_{i-1}^{~} )$,
and the transpose of
${\tilde E}_i^{\rm R}( \sigma'_i \xi'_{i-1} | \sigma_i^{~} \xi_{i-1}^{~} )$. 
The structure of $ {\tilde E}_i^{~} = {\tilde E}_i^{\rm L} \cdot {\tilde E}_i^{\rm R }$
 is understood from Fig.7(c) without difficulty. As we explain in the following, 
$| {\tilde \Phi}_{i}^{~} \rangle$ is easily created 
from $| {\tilde \Psi}_{i - 1}^{~} \rangle$.

It is not required to possess ${\tilde E}_i^{\rm L}$ and  ${\tilde E}_i^{\rm R}$ 
explicitly for the construction of ${\tilde \Phi}_{i}^{~}$ in Eq.(3.15).
Let us substitute the singular value decomposition of ${\tilde \Psi}_{i-1}^{~}$ to the r.h.s of Eq.(3.15).
From the assumed symmetry of ${\tilde \Psi}_{i-1}^{~}$, the decomposition is equivalent to
 the diagonalization of ${\tilde \Psi}_{i-1}^{~}$ as
\begin{equation}
{\tilde \Psi}_{i-1}^{~} = {\hat U}_{i-1}^{~} 
\Omega_{i - 1}^{~} {\hat U}_{i-1}^{\dagger} \, ,
\end{equation}
where $\Omega_{i - 1}^{~}$ is a diagonal matrix, and the
square of $\Omega_{i - 1}^{~}$ represents the eigenvalues of the density matrix
${\tilde \rho}_{i-1}^{\rm R} = {\tilde \Psi}_{i-1}^{\dagger} {\tilde \Psi}_{i-1}^{~} 
= {\hat U}_{i-1}^{~} \left( \Omega_{i - 1}^{~} \right)^2_{~} {\hat U}_{i-1}^{\dagger}$. 
%
%
Substituting Eq.(3.16) to ${\tilde \Psi}_{i-1}^{~}$ in Eq.(3.15), we obtain
\begin{eqnarray}
{\tilde \Phi}_{i }^{~} &=& 
\left( {\tilde E}_i^{\rm L} \right) {\hat U}_{i-1}^{~} 
\Omega_{i - 1}^{~}
{\hat U}_{i-1}^{\dagger} 
\left( {\tilde E}_i^{\rm R} \right)^{\dagger}_{~} \nonumber\\
&=&
\left( {\tilde E}_i^{\rm L}  {\hat U}_{i-1}^{~} \right)
\Omega_{i - 1}^{~}
\left( {\tilde E}_i^{\rm R} {\hat U}_{i-1}^{~} \right)^{\dagger}_{~} \nonumber\\
&=&
{\tilde A}_i^{~} \Omega_{i-1}^{~}  {\tilde A}_i^{\dagger} \, ,
\end{eqnarray}
where ${\tilde A}_i^{~} \equiv  {\tilde E}_i^{\rm R}  {\hat U}_{i-1}^{~}$ and 
we have used the assumed symmetry ${\tilde E}_i^{\rm L} = {\tilde E}_i^{\rm R}$. 
From the recursion relation of ${\tilde E}_i^{\rm R}$ in Eq.(3.10)
the new matrix ${\tilde A}_i^{~}$ satisfies the recursion relation
\begin{eqnarray}
{\tilde A}_i^{~} 
&=&
 {\tilde E}_i^{\rm R} {\hat U}_{i-1}^{~} 
= \left[ {\hat U}_{i-1}^{\dagger} \left( id \cdot {\tilde E}_{i-1}^{\rm R} \right) {\hat U}_{i-2}^{~}  
\right] {\hat U}_{i-1}^{~}
\nonumber\\
&=&
\left[ {\hat U}_{i-1}^{\dagger} \left( id \cdot {\tilde E}_{i-1}^{\rm R} {\hat U}_{i-2}^{~} \right)   
\right] {\hat U}_{i-1}^{~}
\nonumber\\
&=&
\left[ {\hat U}_{i-1}^{\dagger} \left( id \cdot {\tilde A}_{i-1}^{~} \right) \right] {\hat U}_{i-1}^{~}
\, .
\end{eqnarray}
The initial condition of this relation is obtained from the definition of 
${\tilde E}_i^{\rm R}$ in Eq.(3.9). Substituting ${\tilde E}_3^{\rm R} =
{\hat U}_2^{\dagger} I_3^{~} {\hat V}_1^{~}$ to ${\tilde A}_3^{~} = {\tilde E}_3^{\rm R}  {\hat U}_{2}^{~}$
we obtain
${\tilde A}_3^{~} = ( {\hat U}_2^{\dagger} I_3^{~} {\hat V}_1^{~} ) {\hat U}_{2}^{~}$
$= \left[ {\hat U}_2^{\dagger} ( id \cdot  {\hat V}_1^{~} ) \right] {\hat U}_{2}^{~}$, and
thus we find the initial condition
\begin{equation}
{\tilde A}_2^{~} = V_1^{~}.
\end{equation}
It might be helpful to rewrite Eq.(3.18) by the element:
\begin{eqnarray}
&&{\tilde A}_i^{~}( \sigma_i^{~} \xi_{i-1}^{~} | \eta_{i}^{~} ) 
= \sum_{\sigma_{i-1}^{~} \xi_{i-2}^{~}  \eta_{i-1}^{~}}^{~}
U_{i-1}^{~}( \sigma_{i-1}^{~} \xi_{i-2}^{~} | \xi_{i-1}^{~} ) \nonumber\\
&&
\,\,\,\,\,\, \times \, 
{\tilde A}_{i-1}^{~}( \sigma_{i-1}^{~} \xi_{i-2}^{~} | \eta_{i-1}^{~} ) \,
U_{i-1}^{~}( \sigma_i^{~} \eta_{i-1}^{~} | \eta_{i}^{~} ) \, .
\end{eqnarray}
Figure 8 shows the graphical
representation of Eqs.(3.18) and (3.20).

\begin{figure}
\epsfxsize=55mm 
\centerline{\epsffile{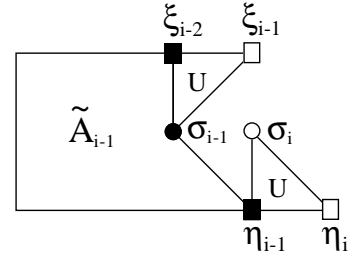}}
\caption{Recursive relations of ${\tilde A}_i^{~}$. 
${\tilde A}_i^{~} = \left[ {\hat U}_{i-1}^{\dagger} \left( id \cdot 
{\tilde A}_{i-1}^{~} \right) \right] {\hat U}_{i-1}^{~}$ for $i \ge 4$ 
in Eqs.(3.18) and (3.20).}
\label{fig:y}
\end{figure}

Using the recursion relation Eq.(3.18) and the approximation from 
Eq.(2.19)' to Eq.(3.13), let us reformulate the recursive creation of
the state vectors proposed in the previous section.
Suppose that we have the renormalized transfer matrix ${\tilde T}_{i-1}^{~}$, 
the state vector $| {\tilde \Psi}_{i-1}^{~} \rangle$
and the matrix ${\tilde A}_{i-1}^{~}$. We obtain 
$| {\tilde \Psi}_{i}^{~} \rangle$ as follows.
\\
\begin{itemize}
\item[(a)] Perform the diagonalization ${\tilde \Psi}_{i-1}^{~}$ 
$\rightarrow {\hat U}_{i-1}^{~} \Omega_{i - 1}^{~}
{\hat U}_{i-1}^{\dagger}$ in Eq.(3.16) to obtain $U_{i-1}^{~}$ 
and $\Omega_{i-1}^{~}$.
\item[(b)] Obtain ${\tilde A}_{i}^{~} = 
\left[ {\hat U}_{i-1}^{\dagger} \left( id \cdot {\tilde A}_{i-1}^{~} \right) \right] {\hat U}_{i-1}^{~}$
by Eq.(3.18).
\item[(c)] Extend the linear dimension of the renormalized transfer matrix by ${\tilde T}_{i}^{\rm R} = 
{\hat U}_{i-1}^{\dagger} \left( W \cdot {\tilde T}_{i-1}^{\rm R} \right)   {\hat U}_{i-1}^{~}$ 
in Eq.(3.3) and the same for ${\tilde T}_{i}^{\rm L}$. 
\item[(d)] Obtain ${\tilde \Phi}_i^{~} =
{\tilde A}_i^{~} \Omega_{i-1}^{~} {\tilde A}_i^{\dagger}$. (Eq.(3.17).)
\item[(e)] Multiply ${\tilde T}_{i}^{~}$ to $| {\tilde \Phi}_i^{~} \rangle$. 
Admitting the approximation from Eq.(2.19)' to Eq.(3.13), substitute
 the obtained vector ${\tilde T}_{i}^{~} | {\tilde \Phi}_i^{~} \rangle$ 
to $| {\tilde \Psi}_i^{~} \rangle$ and proceed to the next iteration.
\end{itemize}
~\\
With use of this recursive calculation, we obtain the state vectors and the 
partition functions explained in the previous section. (The obtained data
is of use for the finite size scaling at the criticality.~\cite{CTMRG2})

The recursive calculation starts from the initial condition given by Eq.(3.19).
The calculation can be stopped when the observed physical 
quantities, such as the free energy per site, converge to their values at the 
large system size limit. 
Alternatively, the difference between ${\tilde \rho}_i^{\rm R} / {\tilde Z}_i^{~}$ and 
 ${\tilde \rho}_{i+1}^{\rm R} / {\tilde Z}_{i + 1}^{~}$ 
 (or in some cases ${\tilde \rho}_i^{\rm R} / {\tilde Z}_i^{~}$ and 
 ${\tilde \rho}_{i+2}^{\rm R} / {\tilde Z}_{i + 2}^{~}$) 
 can be used for the confirmation of the convergence, since
 when $i$ is several times larger than the correlation length the
 recursive procedures (a)-(e) becomes just the repetition of the same
 calculation. It should be noted that even 
 in the large $i$ limit the condition ${\hat U}_{i+1}^{~} =  {\hat U}_{i}^{~}$ is not always 
 satisfied, as reported by Okunishi {\it et al}.~\cite{Okunishi}

 The recursive process (a)-(e) is actually a special  case of the numerical 
algorithm of the 
`product wave function renormalization group (PWFRG) method'.~\cite{PWFRG} 
In order to clarify this point let us consider the numerical problem of obtaining
the largest-eigenvalue eigenvector of the renormalized transfer matrix ${\tilde T}_i^{~}$ 
during the calculation by infinite system DMRG method. Normally one of
the power,
the Lanczos, the Davidson, or the Arnoldi methods is employed for the 
diagonalization of  ${\tilde T}_i^{~}$. The computational 
time required for these diagonalization methods
is greatly dependent on the choice of the initial vector, and therefore it is important to create
a good initial (or trial) vector for the diagonalization of ${\tilde T}_i^{~}$. 
From the arguments in the previous section --- see Eqs.(2.19) and (2.20) ---
it is apparent that 
$| {\tilde \Psi}_i^{~} \rangle = {\tilde P}_i^{~} | {\tilde \Psi}_{i - 1}^{~} \rangle$ 
in Eq.(2.19)' is a candidate of the initial vector. (The choice
is not so bad if the correlation length of the system is 
much shorter than the system size $2i$, as was explained in the
paragraph that contains Eq.(2.4), and
expected to be of use for modest system size.)
Since ${\tilde P}_i^{~}$ is further 
approximated by ${\tilde T}_i^{~} {\tilde E}_i^{~}$, we can say that 
 ${\tilde E}_i^{~}$ has a function of extending the vector dimension
 of $| {\tilde \Psi}_{i - 1}^{~} \rangle$ to create 
 $| {\tilde \Phi}_i^{~} \rangle = {\tilde E}_i^{~} | {\tilde \Psi}_{i - 1}^{~} \rangle$ in Eq.(3.14),
and that ${\tilde T}_i^{~}$ improves  $| {\tilde \Phi}_i^{~} \rangle$ to give 
an approximation for 
$| {\tilde \Psi}_i^{~} \rangle$. We find that 
$| {\tilde \Phi}_{i}^{~} \rangle$ is also of use as the initial vector, 
since the diagonalization methods creates
${\tilde T}_i^{~} | {\tilde \Phi}_{i}^{~} \rangle \sim | {\tilde \Psi}_i^{~} \rangle$, 
$( {\tilde T}_i^{~} )^2_{~} | {\tilde \Phi}_{i}^{~} \rangle
\sim {\tilde T}_i^{~} | {\tilde \Psi}_i^{~} \rangle$, 
$( {\tilde T}_i^{~} )^3_{~} | {\tilde \Phi}_{i}^{~} \rangle
\sim ( {\tilde T}_i^{~} )^2_{~} | {\tilde \Psi}_i^{~} \rangle$, etc., inside their
numerical procedure, and therefore to start the diagonalization 
from $| {\tilde \Phi}_{i}^{~} \rangle$ is almost as efficient as to start from 
$| {\tilde \Psi}_i^{~} \rangle$. 

Based on the above discussion about the initial vector, we find that
the recursion relation Eq.(3.18) can be efficiently used for the numerical
acceleration of the infinite system DMRG algorithm. The way is simply
to replace the step (e) of the recursive calculations (a)-(e) by
\\
\begin{itemize}
\item[(e')] Using $| {\tilde \Phi}_i^{~} \rangle$ as the initial vector for the 
Lanczos  diagonalization (or Davidson, Arnoldi diagonalization or 
the power method)~\cite{power} of ${\tilde T}_{i}^{~}$. 
Substitute the obtained   largest-eigenvalue eigenvector of ${\tilde T}_{i}^{~}$
to $| {\tilde \Psi}_{i}^{~} \rangle$ and proceed to the next iteration.
\end{itemize}
~\\
The modified recursive method, which consists of (a), (b), (c), (d), and (e'),
 is the numerical algorithm of the original version of the 
 PWFRG method.~\cite{PWFRG} So far the initial 
condition of this method --- the matrix element of ${\tilde A}_2^{~}$ ---
was set empirically. What we have clarified here is that the initial condition 
${\tilde A}_2^{~}$ is given by $V_1^{~}$ in Eq.(3.19), which represent
the boundary condition of the system.~\cite{PWFRGX}
One can easily verify that the calculation by (a)-(e') gives the same
result as the infinite system DMRG method, since the DMRG method
consists of (a), (c), and the diagonalization of ${\tilde T}_i^{~}$. 

Even with the use of the initial vector ${\tilde \Phi}_i^{~}$ the diagonalization 
step (e') occasionally consume large amount of computational time. We can 
improve the situation, considering the
half-way between (e) and (e'). For example, if we employ the Lanczos
diagonalization, (e) or (e') can be replaced by
\\
\begin{itemize}
\item[(e'')] Using $| {\tilde \Phi}_i^{~} \rangle$ as the initial vector for the 
Lanczos diagonalization of ${\tilde T}_{i}^{~}$, perform only one or only a few 
Lanczos steps. Substitute the obtained Lanczos vector to
$| {\tilde \Psi}_{i}^{~} \rangle$ and proceed to the next iteration.
\end{itemize}
~\\
This is a realistic choice if one already has a computational program of the
infinite system DMRG method, and if only the physical quantities at the large system
size limit $i \rightarrow \infty$ is required.~\cite{Hieida}

If we store all the block spin transformations ${\hat U}_{2}^{~}$, ${\hat U}_{3}^{~}$, $\ldots$,
${\hat U}_{i-1}^{~}$ obtained successively at the step(a), we
implicitly have the variational MPS~\cite{Ostlund,Ostlund2,Ostlund3,Takasaki,Verstraete}
\begin{equation}
\left[ {\hat U}_{2}^{~} \cdots {\hat U}_{i-1}^{~} \right]  \,\, \Omega_{i - 1}^{~} \,\,
\left[ {\hat U}_{i-1}^{\dagger} \cdots {\hat U}_{2}^{\dagger} \right] \, ,
\end{equation}
which
approximates $| \Psi_{i-1}^{~} \rangle$. The formulation can be obtained
by considering the structure of 
$\langle {\tilde \Psi}_{i-1}^{~} | {\tilde T}_{i-1}^{~} | {\tilde \Psi}_{i-1}^{~} \rangle$
which well approximates 
$\langle \Psi_{i-1}^{~} | T_{i-1}^{~} | \Psi_{i-1}^{~} \rangle$.
The fact enables us to use  the iterative process of the PWFRG
method (a)-(e'), combine with the finite system DMRG method. 
For example, the process  (a)-(e') can be used for the initial  
set-up of the variational MPS, that are further improved by the 
finite system DMRG sweeps. Equally, after finishing
the finite system DMRG calculation for a certain system size $2N$, 
one can switch to the process (a)-(e') again, 
and after $M$ iterations to obtain a good variational MPS for the next finite
system DMRG sweep at  the size $2(N + M)$. The step (e') can be
replaced by (e) or (e''), since the finite system DMRG sweep 
improves the extended MPS. Such a switching process is of use
when precise numerical data calculated by the finite system DMRG method are
required for finite size scaling analyses.

\section{Conclusions and Discussions}

We have explained the physical background of the PWFRG method from the
view point of the MPS formalism. In section II, we first consider the 
calculation of the partition function of the IRF model on a finite size cluster. 
We observe the fact that the largest-eigenvalue eigenvector of a transfer 
matrix of width $2N$ can be approximated by the state vector that
corresponds to lower half (in Fig.2(a)) of the system of width $2N$. 
By use of the fact, we
construct an iterative numerical method to calculate the partition function 
up to arbitrary system size, which is equally of use as the CTMRG 
method.~\cite{CTMRG,CTMRG2} 
In Section III, we analyze the the whole structure of the renormalized 
transfer matrices, which appears in the formalism shown in \S 2.
Observing the matrix product construction of the state vector, we find the
way of increasing the mumber of matrices contained in MPS when we
increment the system size. The extended MPS can be efficiently used 
as the initial vector for the diagonalization of the renormalized transfer matrix,
which appears in the infinite system DMRG method. As a result, we
obtain the numerical algorithm of the PWFRG method. The initial condition 
for the PWFRG method, which has been set empirically, is clearly determined 
by Eq.(3.19) that corresponds to the boundary condition of the system.

The numerical method shown in \S 2 has clear physical meaning as
the partition function calculation of finite size cluster of square shape. The 
same result can be obtained by the PWFRG method with the step (e) 
in the previous section. The PWFRG method with (e') gives exactly the
same result as the infinite system DMRG method, where the 
calculation of the largest-eigenvalue eigenvector of the renormalized
transfer matrix is faster than the conventional infinite system DMRG method.
One can use the PWFRG method with (e'') if the convergence of the
calculated physical quantities toward the large system size limit is
slow with the use of (e) or (e'). The use of (e'') accelerates the convergence
of the free energy per site, but the calculated data before completing 
the convergence do not have clear physical meaning. After the 
convergence, the PWFRG method with (e), (e') and (e'') give the same result about the 
thermodynamic limit of the system.

We comment that the PWFRG method can also be applied to 
1D quantum systems, in addition to 2D classical systems.~\cite{Hieida,Hagiwara,Okunishi,Narumi,Yoshikawa}
The simplest example is a 
uniform spin chain, whose Hamiltonian is written as a sum of position 
independent local terms $h$. In this case we obtain the 
superblock Hamiltonian ${\tilde H}_i^{~} \equiv {\tilde H}_i^{\rm L} + h + {\tilde H}_i^{\rm R}$ 
using the recursive relation
$
{\tilde H}_{i+1}^{\rm R} \equiv {\hat U}_i^{\dagger} ( h +  {\tilde H}_i^{\rm R} ) {\hat U}_{i}^{~}
$
similar to the extension process of the renormalized transfer matrix
 $
{\tilde T}_{i+1}^{\rm R} \equiv {\hat U}_i^{\dagger} ( W \cdot  {\tilde T}_i^{\rm R} ) {\hat U}_{i}^{~}
$. 
Implementing the diagonalization of ${\tilde H}_i^{~}$ into the recursive 
procedure in the PWFRG method (a)-(e') in \S 3, 
one can obtain the ground state property of the quantum 
system in the thermodynamic limit. As for the classical systems,
use of (a)-(e'') is efficient for the acceleration of the convergence
toward the large system size limit.

About the recursive creation of variational state for infinite 1D quantum Hamiltonians, 
the CTMRG method can also be employed for this purpose. Consider
the infinite size chess board system generated by the Trotter 
decomposition~\cite{Trotter,Suzuki} applied to the density matrix 
$\exp( - \beta H )$ at zero temperature.
If we divide the 2D lattice into 4 CTMs by cuts to the diagonal directions, 
the CTM at the top and the bottom corresponds to the variational state, 
and those at the left and the right correspond to the exponential of the corner Hamiltonian.
This approach is quite similar to the system extension method explained in \S 2.
In the limit of the zero imaginary time step in the Trotter decomposition, 
one obtains a DMRG formalism for the corner Hamiltonian. 
The formulation is similar to the recently proposed
renormalization formalism on the corner Hamiltonian by Okunishi,~\cite{okucorn}
but the block spin transformation is obtained from the diagonalization of the
density matrix, instead of the diagonalization of the corner Hamiltonian.

A.G. is supported by the VEGA grant No. 2/3118/23.


\begin{thebibliography}{99}
\bibitem{DMRG} S.~R.~White: Phys. Rev. Lett. {\bf 69} (1992) 2863; 
Phys. Rev. B {\bf 48} (1993) 10345.
\bibitem{DMRG2} {\it Density-Matrix  Renormalization
- A new numerical method in physics -,} eds.
I.~Peschel, X.~Wang, M.~Kaulke and K.~Hallberg, (Springer Berlin, 1999),
and references there in.
\bibitem{DMRG3} U. Schollw\"ock: Rev. Mod. Phys. {\bf 77} (2005) 259.
\bibitem{DMRG4} T.~Nishino: J. Phys. Soc. Jpn. {\bf 64} (1995) 3598.
\bibitem{Fannes} M.~Fannes, B.~Nachtergale  and R.~F.~Werner: Europhys. Lett. 
{\bf 10} (1989) 633: Commun. Math. 
Phys. {\bf 144} (1992)  443: Commun. Math. Phys. {\bf 174}  (1995) 477.
\bibitem{Ostlund} S.~\"Ostlund and S.~Rommer: Phys. Rev. Lett {\bf 75} (1995) 3537.
\bibitem{Ostlund2} S.~Rommer and S.~\"Ostlund: Phys. Rev. B {\bf 55} (1997) 2164.
\bibitem{Ostlund3} M.~Andersson, M.~Boman, and S.~\"Ostlund: Phys. Rev. B {\bf 5}9 (1999) 10493.
\bibitem{Takasaki} H.~Takasaki, T.~Hikihara, and T.~Nishino: J. Phys. Soc. Jpn. {\bf 68} (1999) 1537.
\bibitem{Verstraete} 
F.~Verstraete, J.J.~Garc\'ia-Ripoll, and J.I.~Cirac: Phys. Rev. Lett. {\bf 93} (2004) 207204.
\bibitem{Verstraete2} 
F.~Verstraete, D.~Porras, and J.I.~Cirac: Phys. Rev. Lett. {\bf 93} (2004) 227205.
\bibitem{Verstraete3} V.~Murg, F.~Verstraete, and J.I.~Cirac: cond-mat/0501493.
\bibitem{Acce} S.R.~White and I.~Affleck: Phys. Rev. B {\bf 54} (1996) 9862.
\bibitem{Acce2} S.R.~White: Phys Rev Lett. {\bf 77} (1996) 3633.
\bibitem{PWFRG} T.~Nishino and K.~Okunishi: J. Phys. Soc. Jpn. {\bf 64} (1995) 4084.
\bibitem{Acts} N.~Akutsu and Y.~Akutsu: Phys. Rev. B {\bf 57} (1998) R4233;
N.~Akutsu and Y.~Akutsu: Prog. Theor. Phys. {\bf 105} (2001) 123.
\bibitem{Acts3} 
N.~Akutsu, Y.~Akutsu, and T.~Yamamoto: Prog. Theor. Phys. {\bf 105} (2001) 361;
N.~Akutsu, Y.~Akutsu, and T.~Yamamoto: 
Phys. Rev. B {\bf 64} (2001) 085415;
N.~Akutsu, Y.~Akutsu, and T.~Yamamoto: 
Journal of Crystal Growth {\bf 237-239} (2002) 14;
N.~Akutsu, Y.~Akutsu, and T.~Yamamoto:
Phys. Rev. B {\bf 67} (2003) 125407.
\bibitem{Hieida} Y.~Hieida, K.~Okunishi and Y.~Akutsu:
Phys. Lett. A {\bf 233} (1997) 464.
\bibitem{Hagiwara} 
M.~Hagiwara, Y.~Narumi, K.~Kindo, M.~Kohno, H.~Nakano, R.~Sato, and M.~Takahashi: 
Phys. Rev. Lett. {\bf 80} (1998) 1312.
\bibitem{Okunishi} K.~Okunishi, Y.~Hieida, and Y.~Akutsu: Phys. Rev. B {\bf 59}
(1999) 6806;
K.~Okunishi, Y.~Hieida, and Y.~Akutsu
Phys. Rev. E {\bf 59} (1999) R6227;
Y.~Hieida, K.~Okunishi, and Y.~Akutsu:
New Journal of Physics {\bf 1} (1999) 7.1;
K.~Okunishi, Y.~Hieida, and Y.~Akutsu: 
Phys. Rev. B {\bf 60} (1999) R6953;
Y.~Hieida, K.~Okunishi, and Y.~Akutsu: 
Phys. Rev. B {\bf 64} (2001) 224422.
\bibitem{Narumi} Y.~Narumi, K.~Kindo, M.~Hagiwara, H.~Nakano, A.~Kawaguchi
K.~Okunishi, and M.~Kohno: 
Phys. Rev. B {\bf 69} (2004) 174405.
\bibitem{Yoshikawa} S.~Yoshikawa, K.~Okunishi, M.~Senda and S.~Miyashita: 
J. Phys. Soc. Jpn. {\bf 73} (2004) 1798.
\bibitem{Baxter} R.J.~Baxter: 
{\it Exactly Solved Models in Statistical Mechanics} (Academic Press, London, 1982).
\bibitem{CTMRG} T.~Nishino and K.~Okunishi: 
J. Phys. Soc. Jpn. {\bf 65} (1996) 891; 
J. Phys. Soc. Jpn. {\bf 66} (1997) 3040.
\bibitem{CTMRG2} T.~Nishino, K.~Okunishi, and M.~Kikuchi: Phys. Lett. A {\bf 213}
 (1996) 69.
\bibitem{general} The first condition can be easily removed, just by treating the left
half of the system independently from the right half. To remove the second condition, 
one has to diagonalize the asymmetric density matrix to obtain bi-orthogonal 
block spin transformations.
\bibitem{cluster} The shape 
of the cluster in Fig.1 is close to that was treated by Baxter in his method of CTM.~\cite{Baxter} 
\bibitem{power} The Lanczos, the Davidson, and the Arnoldi diagonalizations
can be interpreted as improvements of the power method.
\bibitem{PWFRGX} We list the correspondence of the procedures (a)-(e')  to equations in 
Ref.[15]: (a) $\rightarrow$ Eq.(5),\cite{PWFRG} (b) $\rightarrow$ Eq.(7),\cite{PWFRG} (c) $\rightarrow$ Eq.(6),\cite{PWFRG} (d) $\rightarrow$ Eq.(3),\cite{PWFRG} 
(e') $\rightarrow$ Eq.(4).\cite{PWFRG}
\bibitem{Trotter} H.~F.~Trotter: Proc. Am. Math. Soc. {\bf 10}
(1959) 545.  
\bibitem{Suzuki} M.~Suzuki: Prog. Theor. Phys. {\bf 56} (1976) 1454.  
\bibitem{okucorn} K.~Okunishi: preprint, cond-mat/0507195.
\end{thebibliography}
\end{document}